\documentclass[iop]{emulateapj}

\usepackage{natbib}
\begin{document}
\submitted{Submitted to ApJ 2015-01-21, under review}
\title{The Distance to Nova V959 Mon from VLA Imaging}
\author{J.~D.~Linford\altaffilmark{1}, V.~A.~R.~M.~Ribeiro\altaffilmark{2,3},  L.~Chomiuk\altaffilmark{1}, T.~Nelson\altaffilmark{4}, J.~L.~Sokoloski\altaffilmark{5}, M.~P.~Rupen\altaffilmark{6}, K.~Mukai\altaffilmark{7,8}, T.~J.~O'Brien\altaffilmark{9}, A.~J.~Mioduszewski\altaffilmark{10}, and J.~Weston\altaffilmark{5}}

\altaffiltext{1}{Department of Physics and Astronomy, Michigan State University, East Lansing, MI 48824, USA}
\altaffiltext{2}{Department of Astrophysics/IMAPP, Radboud University, PO Box 9010, 6500 GL, Nijmegen, The Netherlands}
\altaffiltext{3}{Astrophysics, Cosmology, and Gravity Centre, Department of Astronomy, University of Cape Town, Private Bag X3, Rondebosch 7701, South Africa}
\altaffiltext{4}{School of Physics and Astronomy, University of Minnesota, 116 Church St SE, Minneapolis, MN 55455}
\altaffiltext{5}{Columbia Astrophysics Laboratory, Columbia University, New York, NY, USA}
\altaffiltext{6}{Herzberg Institute of Astrophysics, National Research Council of Canada, Penticton, BC, Canada}
\altaffiltext{7}{Center for Space Science and Technology, University of Maryland Baltimore County, Baltimore, MD 21250, USA}
\altaffiltext{8}{CRESST and X-ray Astrophysics Laboratory, NASA/GSFC, Greenbelt MD 20771 USA}
\altaffiltext{9}{Jodrell Bank Centre for Astrophysics, Alan Turing Building, University of Manchester, Manchester, M13 9PL, UK}
\altaffiltext{10}{National Radio Astronomy Observatory, P.O. Box 0, Socorro, NM 87801, USA}
\email{jlinford@.msu.edu}

\begin{abstract}
Determining reliable distances to classical novae is a challenging but crucial step in deriving their ejected masses and explosion energetics.  Here we combine radio expansion measurements from the Karl G. Jansky Very Large Array with velocities derived from optical spectra to estimate an expansion parallax for nova V959 Mon, the first nova discovered through its $\gamma$-ray emission.  We spatially resolve the nova at frequencies of 4.5--36.5 GHz in nine different imaging epochs.  The first five epochs cover the expansion of the ejecta from 2012 October to 2013 January, while the final four epochs span 2014 February to 2014 May.  These observations correspond to days 126 through 199 and days 615 through 703 after the first detection of the nova.  The images clearly show a non-spherical ejecta geometry.  Utilizing ejecta velocities derived from 3D modelling of optical spectroscopy, the radio expansion implies a distance between $0.9\pm0.2$ and $2.2\pm0.4$ kpc, with a most probable distance of $1.4\pm0.4$ kpc.  This distance implies a $\gamma$-ray luminosity much less than the prototype $\gamma$-ray-detected nova, V407 Cyg, possibly due to the lack of a red giant companion in the V959 Mon system.  V959 Mon also has a much lower $\gamma$-ray luminosity than other classical novae detected in $\gamma$-rays to date, indicating a range of at least a factor of 10 in the $\gamma$-ray luminosities for these explosions.
\end{abstract}

\section{Introduction}
\label{intro}

The discovery of a $\gamma$-ray transient coincident with the 2010 nova event in V407 Cyg was a source of much excitement and surprise for the nova and $\gamma$-ray communities alike \citep{Abdo10}.  Although MeV $\gamma$-rays produced by nuclear decay in nova ejecta have been predicted for many years \citep[e.g., ][and references therein]{Hernanz13}, GeV emission from novae, as detected with {\it Fermi} Large Area Telescope (LAT; Atwood et al. 2009), had received little attention prior to the event in V407 Cyg \citep[with the notable exception of ][]{Tatischeff07}. Abdo et al. (2010) suggested that a blast wave driven into the wind from the Mira giant companion accelerated particles to relativistic speeds and produced $\gamma$-rays through either leptonic or hadronic secondary interaction. In follow-up work exploring the X-ray properties of V407 Cyg, \citet{Nelson12} also claimed that the presence of a red giant companion was the primary characteristic of the system responsible for generating such a $\gamma$-ray event and efficient acceleration of particles. They predicted that $\gamma$-ray emission from novae would be very rare, as most systems have main sequence (MS) donors \citep[see also][]{Lu11}. 

It did not take long for nature to prove this prediction incorrect. In 2012, the {\it Fermi}-LAT detected two new transients that were spatially coincident with novae. The association of Fermi J1750-3243 with the nova V1324 Sco was made within a few weeks of the outburst \citep{Cheung12a}. However, the nature of $\gamma$-ray transient FGL J0639+0548 remained a mystery for several months after discovery \citep{Cheung12b}, because its proximity to the sun prevented follow-up by optical observers. When the region became optically observable a few months after $\gamma$-ray detection, a nova was discovered at the location of the {\it Fermi} transient \citep{Cheung12c}. Thus V959 Mon was the first nova \textit{discovered} in the $\gamma$-rays.  In 2013, the two naked-eye visible novae V339 Del and V1369 Cen joined the collection of novae detected in $\gamma$-rays (Hays et al. 2013, Cheung et al. 2013, Ackermann et al. 2014).
\begin{center}
\begin{figure*}[t]
\includegraphics[width=7.0in]{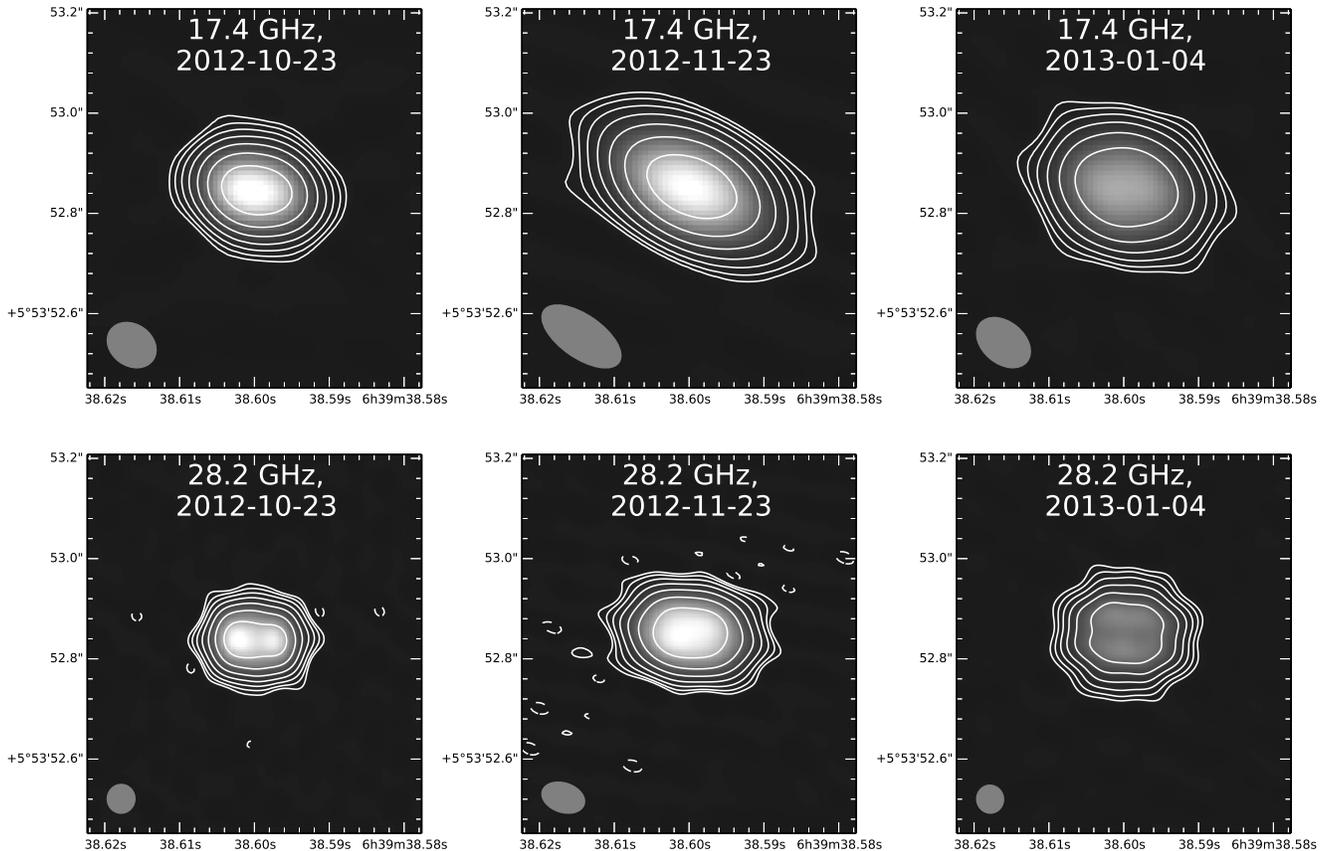}
\vspace{-0.45cm}
\caption{Selected images of nova V959 Mon over the course of the 2012-2013 VLA A-configuration campaign.  Left-hand column: 126 days from \textit{Fermi} detection. Middle column: 157 days after \textit{Fermi} detection.  Right-hand column: 199 days after \textit{Fermi} detection.  Contour levels are -0.0004, 0.0004, 0.0008, 0.0016, 0.0032, 0.0064, 0.0128, and 0.0256 Jy/beam in all images.  The field of view for all image is 0.66 arcsec in RA by 0.954 arcsec in DEC. The synthesized beams are shown by the gray ellipse in the bottom left of each image.}
\label{Aconfig1}
\end{figure*}
\end{center}
The novae V1324 Sco, V959 Mon, V339 Del, and V1369 Cen do not show evidence of an evolved companion \citep{Greimel12,Wagner12,Darnley13,Schaefer14}, making them very different from V407 Cyg. Greimel et al.~discuss the identification of the likely progenitor of V959 Mon in images obtained as part of the IPHAS survey \citep{Drew05}.  They find only a faint source ($r \approx 17.9$ mag) at the location of the nova, which is too faint to be associated with a giant star within the Milky Way. Furthermore, no spectral signatures of an evolved companion (e.g. molecular bands of TiO and VO) were observed in the spectra obtained during outburst.  X-ray observations (e.g., Orio \& Pa 2012, Osborne et al. 2013) indicated the companion star was partially eclipsing and had an orbital period of 7.1 hours.  The 7.1 hour periodicity was also observed at optical wavelengths (e.g., Wagner et al. 2013, Hambsch et al. 2013, Munari et al. 2013).  Page et al. (2013) observed V959 Mon with X-ray, ultraviolet, and near-infrared instruments and concluded that the companion star was a near-main-sequence star.  

In fact, other than the $\gamma$-ray detections themselves, no property of V1324 Sco, V339 Del, V1369 Cen, or V959 Mon is particularly unusual compared to other novae, although V339 Del and V1369 Cen were quite bright in the optical.  A large number of other novae have been optically discovered since the launch of \emph{Fermi}, yet none of these have been detected as a $\gamma$-ray source \citep{LAT14}.  An obvious question therefore arises---{\it why did V959 Mon, V1324 Sco, and V339 Del produce bright $\gamma$-ray emission while others did not?}  

One possibility, as suggested by \cite{LAT14}, is that V959 Mon and other $\gamma$-ray-detected novae are unusually nearby compared with the general population of novae. Accurate distances to novae are notoriously difficult to determine.  The velocity of absorption systems in the interstellar medium can be measured from low-excitation lines like the \ion{Na}{1}~D doublet, and can be used to place limits on the distance to the nova; however, this method has yet to be extensively tested and relies on knowledge of gas motions in the Galaxy.  In the case of V959 Mon, \ion{Na}{1}~D absorption \citep[e.g,][]{MZ97} yields a distance estimate of $1\lesssim d \lesssim 2$ kpc \citep{Munari13}. Shore et al. (2013) put V959 Mon at a distance of 3.6 kpc by comparing its UV spectra to that of V1974 Cyg.  Santangelo et al. (2013) used the maximum magnitude/rate of decline (MMRD) relationship (e.g., Downes \& Duerbeck 2000) to get a distance of 3.5 kpc.  However, the MMRD relation is plagued by significant uncertainty \citep[e.g.,][]{Kasliwal11, Cao12}, and at any rate requires observations of the true optical peak (certainly not measured for V959 Mon). Both the UV spectroscopic and MMRD method rely on several assumptions about the nova and are subject to significant uncertainties.

The distance to a nova can be directly determined if the ejecta are spatially resolved while still optically thick, and the measured angular expansion rate is compared with the physical expansion velocity (e.g., Woudt et al. 2009).  The velocity is typically derived from measurements of the width of optical emission lines (e.g., Slavin et al. 1995), although infrared forbidden lines have occasionally been used (e.g., Lyke \& Campbell 2009). In the optical, the expansion parallax measurements are difficult because the morphology of the ejecta is not always well-understood \citep{Wade2000} and novae rapidly become optically thin at optical wavelenghts (e.g., Downes \& Duerbeck 2000). Novae remain optically thick at radio wavelengths for much longer---sometimes months. Radio emission in novae is produced primarily by thermal free-free emission from the ionized ejecta \citep{Seaquist_Bode08}. The ejecta can be resolved with radio interferometers over time baselines which are sufficient to measure and confirm the expansion rate. This expansion parallax technique has been used at radio wavelengths to determine distances to classical novae QU~Vul \citep{Taylor_etal87} and V407~Cas \citep{Eyres_etal00}, recurrent nova RS~Oph \citep{Rupen_etal08}, and symbiotic nova HM Sagittae \citep{Eyres_etal95}. 
\begin{center}
\begin{deluxetable*}{cccccccc}[th]
\tablewidth{0 pt}
\tabletypesize{\footnotesize}
\setlength{\tabcolsep}{0.025in}
\tablecaption{ \label{obs}
Parameters of 2012-2013 Radio Images of V959 Mon}
\tablehead{UT Date & $\Delta$t\tablenotemark{a}  & Freq & Synth.~Beam\tablenotemark{b} & $S_{\nu}$\tablenotemark{c} & Major Axis\tablenotemark{d} & Minor Axis\tablenotemark{d} & PA\tablenotemark{e} \\
& (day) & (GHz) & (arcsec$\times$arcsec, degrees) & (mJy) & (arcsec) & (arcsec) & (degrees) }
\startdata
                                  
2012 Oct 23.4 & 126 & 13.54 & $0.16091\times0.13716, +13.36$ & $71.22\pm0.05$ & $0.10641\pm0.00019$ & $0.06028\pm0.00032$ & $+85.48\pm0.21$ \\
 & & 17.46 & $0.12710\times0.10739, +23.49$ & $90.05\pm0.06$ & $0.10752\pm0.00014$ & $0.06173\pm0.00020$ & $+86.87\pm0.16$ \\
 & & 28.21 & $0.08788\times0.06756, -13.59$ & $139.67\pm0.18$ & $0.10806\pm0.00017$ & $0.06531\pm0.00021$ & $+85.25\pm0.18$ \\
 & & 36.53 & $0.06459\times0.05314, -35.20$ & $175.39\pm0.22$ & $0.10676\pm0.00015$ & $0.06619\pm0.00014$ & $+84.20\pm0.15$ \\
\\
2012 Nov 12.3 & 146 & 13.54 & $0.16032\times0.15098, +28.65$ & $69.23\pm0.04$ & $0.12570\pm0.00015$ & $0.07736\pm0.00019$ & $+85.86\pm0.15$ \\
 & & 17.45 & $0.12403\times0.11877, +15.09$ & $84.88\pm0.05$ & $0.12565\pm0.00011$ & $0.07674\pm0.00012$ & $+86.76\pm0.09$ \\
 & & 28.21 & $0.08721\times0.07015, -24.45$ & $108.61\pm0.14$ & $0.12250\pm0.00019$ & $0.07789\pm0.00021$ & $+86.34\pm0.19$ \\
 & & 36.53 & $0.06788\times0.05460, -51.50$ & $130.79\pm0.17$ & $0.11871\pm0.00017$ & $0.07981\pm0.00016$ & $+86.77\pm0.18$ \\
\\
2012 Nov 23.5 & 157 & 13.54 & $0.29344\times0.14506, +50.56$ & $63.60\pm0.08$ & $0.12773\pm0.00066$ & $0.08305\pm0.00082$ & $+89.99\pm0.67$ \\
 & & 17.45 & $0.23724\times0.11184, +49.59$ &  $77.11\pm0.10$ & $0.12677\pm0.00052$ & $0.08256\pm0.00066$ & $+90.55\pm0.54$ \\
 & & 28.21 & $0.11706\times0.07698, +52.37$ & $126.48\pm0.13$ & $0.12419\pm0.00022$ & $0.08427\pm0.00022$ & $+88.46\pm0.25$ \\
 & & 36.53 & $0.09361\times0.05531, +59.52$ & $152.99\pm0.22$ & $0.12016\pm0.00024$ & $0.08610\pm0.00022$ & $+88.71\pm0.31$ \\
\\
2012 Dec 7.2 & 171 & 13.54 & $0.25068\times0.15713, -60.13$ & $65.36\pm0.04$ & $0.13788\pm0.00032$ & $0.09083\pm0.00036$ & $+85.48\pm0.31$ \\ 
 & & 17.45 & $0.19889\times0.12303, -58.75$ & $81.25\pm0.06$ & $0.13532\pm0.00024$ & $0.09013\pm0.00027$ & $+85.61\pm0.25$ \\
 & & 28.21 & $0.13661\times0.07350, -50.08$ & $137.77\pm0.16$ & $0.12915\pm0.00028$ & $0.09210\pm0.00032$ & $+85.65\pm0.35$ \\
 & & 36.53 & $0.10957\times0.05227, -51.88$ & $172.27\pm0.25$ & $0.12533\pm0.00031$ & $0.09372\pm0.00032$ & $+84.96\pm0.44$ \\
\\
2013 Jan 4.3 & 199 & 13.54 & $0.18791\times0.14346, +30.64$ & $62.61\pm0.04$ & $0.15483\pm0.00022$ & $0.10724\pm0.00030$ & $+88.36\pm0.23$ \\
 & & 17.45 & $0.14803\times0.11436, +33.31$ & $78.46\pm0.06$ & $0.14988\pm0.00021$ & $0.10751\pm0.00025$ & $+89.05\pm0.25$ \\
 & & 28.21 & $0.08154\times0.06572, -3.92$ & $135.25\pm0.20$ & $0.14075\pm0.00021$ & $0.11333\pm0.00021$ & $+87.88\pm0.31$ \\
 & & 36.53 & $0.06100\times0.05088, -9.49$ & $163.29\pm0.33$ & $0.13365\pm0.00029$ & $0.11572\pm0.00028$ & $+88.60\pm0.67$ 
\enddata
\tablenotetext{a}{Time since \emph{Fermi} discovery (2012 June 19). $^{\rm b}$Dimensions of the image's synthesized beam, listing major axis (FWHM in arcsec), minor axis (FWHM in arcsec), and position angle (degrees east of north). $^{\rm c}$ Uncertainties are the formal errors from the Gaussian fit and do not take into account the overly simplistic nature of the single component model. $^{\rm d}$FWHM of fitted Gaussian, after it has been deconvolved from the synthesized beam. $^{\rm e}$Position angle of fitted Gaussian measured in degrees east of north.}
\end{deluxetable*} 
\end{center}
An additional complication in determining the expansion parallax is introduced when the ejecta is not spherically symmetric.  While the radio light curves of many novae have been shown to agree well with an expanding spherical shell (e.g., Hjellming 1996), it is well-known that many novae are not so simple (e.g., Woudt et al. 2009; Chomiuk et al. 2014b).  Recent development in modelling have resulted in new non-spherical models for the novae ejecta (e.g., Ribeiro et al. 2013; Ribeiro et al. 2014).  These non-spherical models allow for a more complicated velocity profile and can be compared with observations to determine the correct velocity to use for the expansion parallax calculation.

A subset of the resolved radio images of V959 Mon obtained with the Karl G.~Jansky Very Large Array (VLA) images were first presented in \cite{Chomiuk14b}.  The nova shows evidence of two components: a denser, slower moving torus of material in the the orbital plane of the binary, and a faster and more diffuse outflow with a bipolar shape.  The $\gamma$-rays are thought to originate from the interaction of the two flows \citep{Chomiuk14b}.  

In this paper, we present observational data from the VLA A-configuration campaigns in 2012 and 2013.  We use the images to derive an expansion parallax distance to the source.  We present our observations and data analysis techniques in Section 2.  In Section 3, we discuss the models used to determine the ejecta velocity profile and compare them with observations.  In Section 4, we use the evolution of the radial extent of the radio emitting ejecta to derive a distance to the source.  We discuss the observed properties of the source, and the implication of our derived distance on the $\gamma$-ray luminosity, in Section 5.  Final conclusions are presented in Section 6.  Throughout the paper, we assume that the start of the outburst $t_{0} = 56097$ (MJD), i.e. the date of the first detection of $\gamma$-rays associated with the source \citep{LAT14}.

\section{Observations and Data Reduction}

VLA radio observations of V959 Mon began soon after the discovery of the $\gamma$-ray transient, as part of a program to follow up on new \textit{Fermi}-LAT transients (NRAO Program S4322).  We then began observing the source after the discovery of its optical counterpart in 2012 August as part of the E-Nova Project (NRAO Program 12B-375).  In this article, we concentrate on the high-frequency data obtained while the VLA was operating in its extended A configuration during two time periods: 2012 October to 2013 January and 2014 February to 2014 May.  A more comprehensive discussion of the entire observing campaign, including analysis of the complete light curve and spectral evolution of V959 Mon, is given in \cite{Chomiuk14b}.  

We obtained images of V959 Mon in Ku (12--18 GHz) and Ka (26.5--40 GHz) bands in seven A-configuration epochs. We also obtained images in C-band (4--8 GHz) in all A-configuration epochs, but the nova was resolved in only the final 2 epochs.  The data were acquired with 2 GHz of bandwidth split between two independently tunable 1-GHz-wide basebands. 

The Ku- and Ka-band exposure times were $\sim$10--15 min on source in each receiver band, but V959 Mon exhibited such high flux densities that excellent signal-to-noise ratios were obtained in this time during the first 5 epochs (Table~\ref{obs}). We used the source J0643+0857 as the gain calibrator and carried out absolute flux density and bandpass calibration using 3C147. Referenced pointing scans were used to ensure accurate pointing; pointing solutions were obtained on both the flux calibrator and gain calibrator, and the pointing solution from the gain calibrator was subsequently applied to our observations of V959 Mon. Fast switching was used between the gain calibrator and V959 Mon with a cycle time of 1.5--2 minutes.  

The C-band exposure times for the final two epochs were $\sim$ 22 minutes on source.  The source J0632+1022 was used as the gain calibrator, while 3C147 was again used for absolute flux density and bandpass calibration.

Data reduction for all epochs and frequencies was carried out using standard routines in AIPS\footnote{http://www.aips.nrao.edu}. The calibrated data were split into their two basebands, thereby providing two frequency points per receiver band. Images were made using Difmap (Shepard 1997), an imaging software package specifically for radio interferometry.  V959 Mon was bright enough during the 2012-2013 set of A configuration observations for full phase and amplitude self-calibration to be carried out.  During the 2014 set of observations, the nova had become so dim that only phase self-calibration was possible. 

During the 2012-2013 campaign, V959 Mon was clearly resolved at 13.5 to 36.5 GHz (see Figure \ref{Aconfig1}), with 1--3 synthesized beams across it (depending on the frequency and epoch). V959 Mon was imaged at significance levels of $>$100$\sigma$ per synthesized beam. The position angles of the fitted Gaussian were consistent with one another and distinct from the position angle of the synthesized beams. In the 2012-10-23 (day 126) observations, the ejecta displayed very similar deconvolved sizes at the four higher frequencies (13.5 GHz to 36.5 GHz), despite very different native resolutions. Later on, the major axis measurements diverged across frequencies in a systematic way (see Section 4 for further discussion), while the minor axis measurements continued to be consistent at all four frequencies.

For the 2012-2013 observations, the flux density of V959 Mon was measured by fitting a 2-D Gaussian to the imaged source with the task \verb|JMFIT| in AIPS. The width and position angle of the Gaussian is allowed to vary, and the angular dimension of V959 Mon are found by deconvolving the synthesized beam from the fitted Gaussian. We also record the integrated flux density of the Gaussian, and the flux density error from \verb|JMFIT|. Our results are presented in Table~\ref{obs}.  The 2014 observations were too complex to be fit by a single component.
\begin{figure}[b]
\includegraphics[width=3.5in]{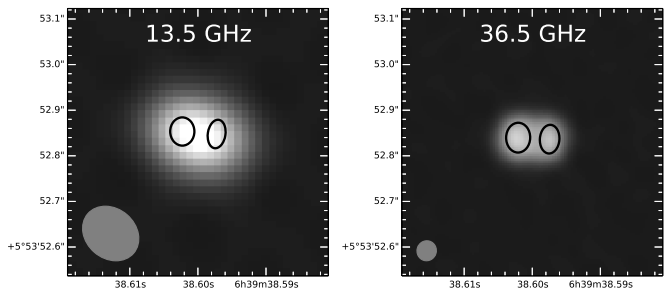}
\caption{Images of nova V959 Mon from 2012-10-23, 126 days after detection.  The elliptical Gaussian models from the \textit{(u,v)} plane are shown in black.  Images are 0.57 arcsec square.}
\label{difmap_mod}
\end{figure}
Single 2-D Gaussian fits provide simple, first-order estimates of the changing dimensions of V959 Mon, although the source is not perfectly described with this profile form.  To better model a bipolar structure, we used Difmap to model the source in the \textit{(u,v)} plane using two elliptical Gaussian components.  Modelling the source in the \textit{(u,v)} plane is often preferable to modelling in the image plane because the \textit{(u,v)} data are not convolved with the restoring beam and are not influenced by the choice of weighting (robust, natural, uniform, etc.) for producing an image.  These two-component models produced better (lower reduced chi-square and smaller rms in the image) fits to the data than single-component models in all of the first five epochs.  The Difmap modelfit procedure consistently placed the two components at either side of the nova in the east-west direction.  This gives us two points per observation to with which to measure the north-south extent of the nova ejecta (see Figure \ref{difmap_mod}).  For the remainder of the paper, we refer to these two-component 2-D Gaussian models as the ``Difmap models''.

\begin{figure}[t]
\includegraphics[width=3.5in]{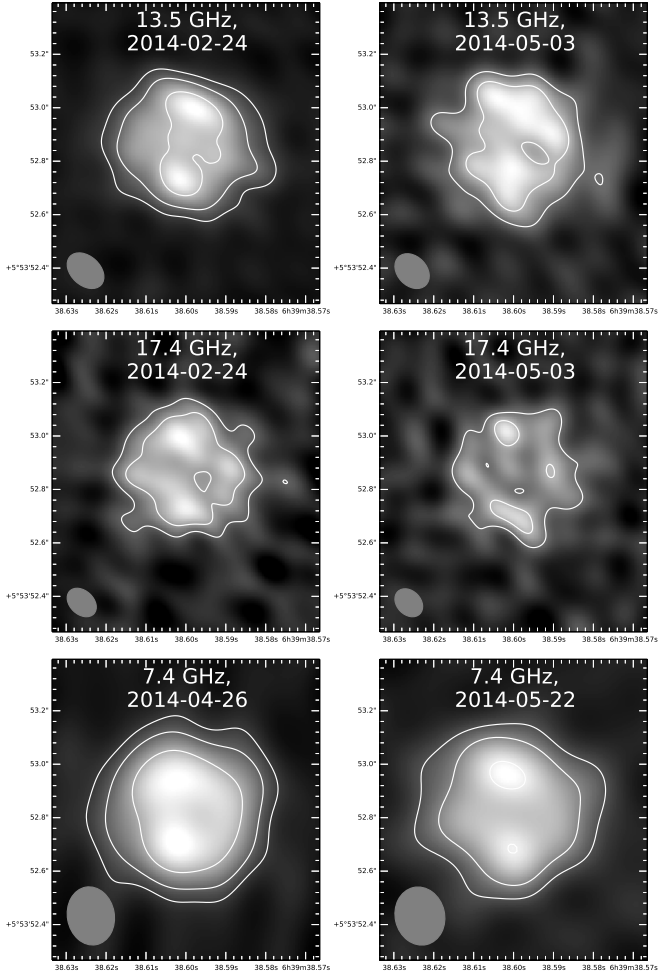}
\caption{Selected images of nova V959 Mon over the course of the 2014 VLA A-configuration campaign.  2014-02-24 is 615 days after \textit{Fermi} detection, 2014-04-26 is 676 days, 2014-05-03 is 683 days, and 2014-05-22 is 703 days.  Contour levels are -0.0001, 0.0001, 0.0002, 0.0004, 0.0008, 0.0016, 0.0032, 0.0064, 0.0128, and 0.0256 Jy/beam in all images.  All images are 1 arcsec square.  The restoring beam is shown by the gray ellipse in the lower left of each image.}
\label{Aconfig2}
\end{figure}

The 2014 A-configuration campaign revealed stark changes in the morphology of the nova ejecta.  The nova went from being brightest along the east-west axis to being brightest along the north-south axis (see Figure \ref{Aconfig2}).  Also, the source had become so diffuse that it was resolved out at Ka-band (28.2 and 36.5 GHz).  Applying a taper to the Ka-band data produced images similar to those at Ku-band (13.5 and 17.4 GHz).  While the Ka-band observations were resolved out, the C-band (7.4 GHz) images were finally nicely resolved.
\section{SHAPE Models of 2012-2013 Data}

Various studies found evidence for a bipolar morphology in V959 Mon (e.g., O'Brien et al. 2012, Shore et al. 2013).  Ribeiro et al. (2013) applied SHAPE\footnote{Available from \url{http://www.astrosen.unam.mx/shape}} (Version 5; Steffen et al. 2011) to replicate the [\ion{O}{3}] emission lines from V959 Mon in order to determine the morphology and system parameters.  At optical wavelengths, the SHAPE models were applied assuming an optically thin morphology, appropriate for the [\ion{O}{3}] emission lines.  They determined a maximum ejecta velocity of $2400^{+300}_{-200}$ km s$^{-1}$, an inclination angle of $82^{\circ}$ and a ``squeeze'' (1 minus the ratio of the minimum extent to maximum extent) of 0.8.  In other words, the largest extent of the nova ejecta is 5 times the size of the minimum extent.  Also from Ribeiro et al. (2013), the maximum extent along the north-south axis was assumed to be one-half the extent along the east-west axis.  Shore et al. (2013) predicted a similar morphology based on their spectroscopic modelling. 

For this paper, we created models of the radio emission in SHAPE using the values determined from Ribeiro et al. (2013) as input.  The key difference between the optical SHAPE and radio SHAPE models is that in the latter case we have now incorporated emission and absorption coefficients for the free-free thermal process.  We calculate the radiative transfer on a 3-D grid incorporating the structure via ray tracing to the observer.  The emission and absorption coefficients are dependent on several parameters such as the ejecta density and temperature, as well as the observing frequency.  Synthetic radio images were generated assuming an emissivity proportional to the density squared.  The density was taken to vary as $r^{-2}$ and the shell was assumed to maintain a constant fractional thickness with the inner radius 0.25$\times$ the outer radius.  The velocity structure is taken as a ``Hubble flow'', where velocity increases linearly with radius, reaching a maximum of 2400 km s$^{-1}$.  The temperature of the ejecta was assumed constant at $65,000$K, the brightness temperature determined from the first VLA Ku-band A-configuration imaging using the standard technique.  This was the earliest resolved image of the nova ejecta, and we assume it was still optically thick at this time and frequency.  

\begin{figure}[t]
\includegraphics[width=3.5in]{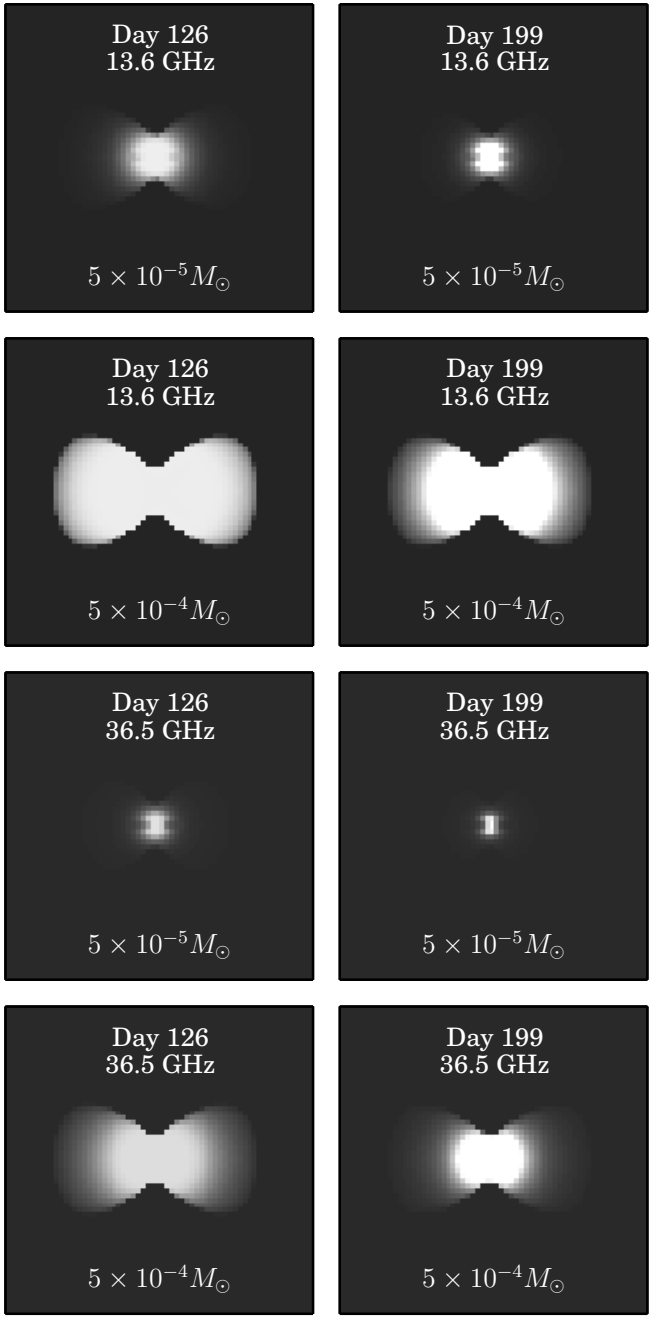}
\caption{Examples of the SHAPE models for V959 Mon.  The field of view is the same for all images.  The flux scales are the same for a given frequency with white indicating higher radio flux density.}
\label{shape_mods}
\end{figure}

The ejecta was modelled on days 126, 157, and 199 after explosion, and at frequencies of 13.6 and 36.5 GHz.  Models were created with ejecta masses of $1\times10^{-5}M_{\odot}$, $5\times10^{-5} M_{\odot}$, $1\times10^{-4}M_{\odot}$, $5\times10^{-4} M_{\odot}$, and $1\times10^{-3} M_{\odot}$.  Models with higher ejecta masses remained optically thick for longer.  The ejecta was assumed to have a filling factor of 1.  The models initially assumed a distance of 1 kpc to match the minimum expected distance from Munari et al. (2012).  Later, the models were adapted in both pixel scale and flux density to correspond to a distance of 2 kpc, the maximum expected distance from Munari et al. (2012).  See Figure~\ref{shape_mods} for examples of the SHAPE models.
It must be noted that the optical spectroscopic observations used to determine the morphology in \cite{Ribeiro13} were made on 2012 October 29, 133\footnote{Ribeiro et al. (2013) reported this observation as day 130 using $t_{0}=56100$ (MJD), but subsequent inspection of the \textit{Fermi} data led to the initial detection date being moved to 3 days earlier on $56097$ (MJD).} days after \textit{Fermi} detection.  This is close to our first high-resolution VLA observation on 2012 October 23, but not simultaneous.  The maximum ejecta velocity was determined from 8 optical spectroscopic observations between day 63 and day 193, again not exactly simultaneous with VLA observations but close.  In principle, the morphology of the ejecta should be set once the material has cleared the orbit of the secondary star, and it should expand at a roughly constant rate from that point.  There is no strong evidence for a deceleration in the [\ion{O}{3}] spectra, especially after day 133 (Ribeiro et al. 2013).  However, there is the possibility that the ejecta underwent some deceleration which we have not taken into account.  Furthermore, the [\ion{O}{3}] observed in the optical is not necessarily the material emitting at radio wavelengths and traces a relatively low density outer region of the ejecta (Ribeiro et al. 2013).  Therefore, careful comparisons between the radio SHAPE models built here and the radio observations are necessary to ensure that the two agree reasonably well.

\begin{center}
\begin{figure*}[t]
\includegraphics[width=7.0in]{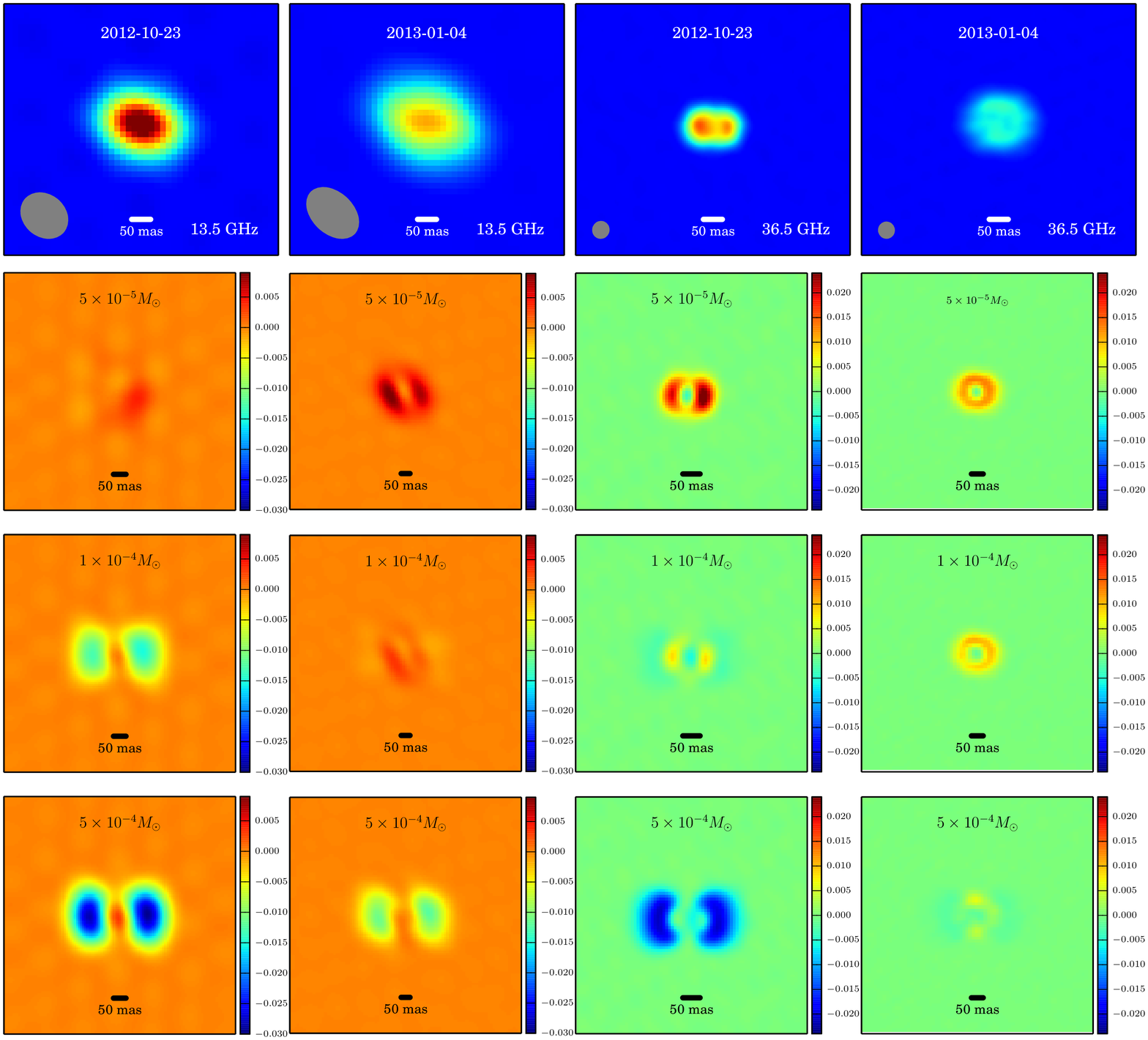}
\caption{Residual images (observed - model at 1 kpc) from 2012-10-23 (day 126) and 2013-01-04 (day 199).  Top row: VLA observation, with gray ellipse to indicate restoring beam.  Second row: Residual, SHAPE model $5\times10^{-5}M_{\odot}$.  Third row: $10^{-4}M_{\odot}$.  Bottom row: $5\times10^{-4}M_{\odot}$. A scalebar in the bottom center of each image indicate an angular size of 50 milliarcseconds.}
\label{res1kpc}
\end{figure*}
\end{center}

\begin{center}
\begin{figure*}[t]
\includegraphics[width=7.0in]{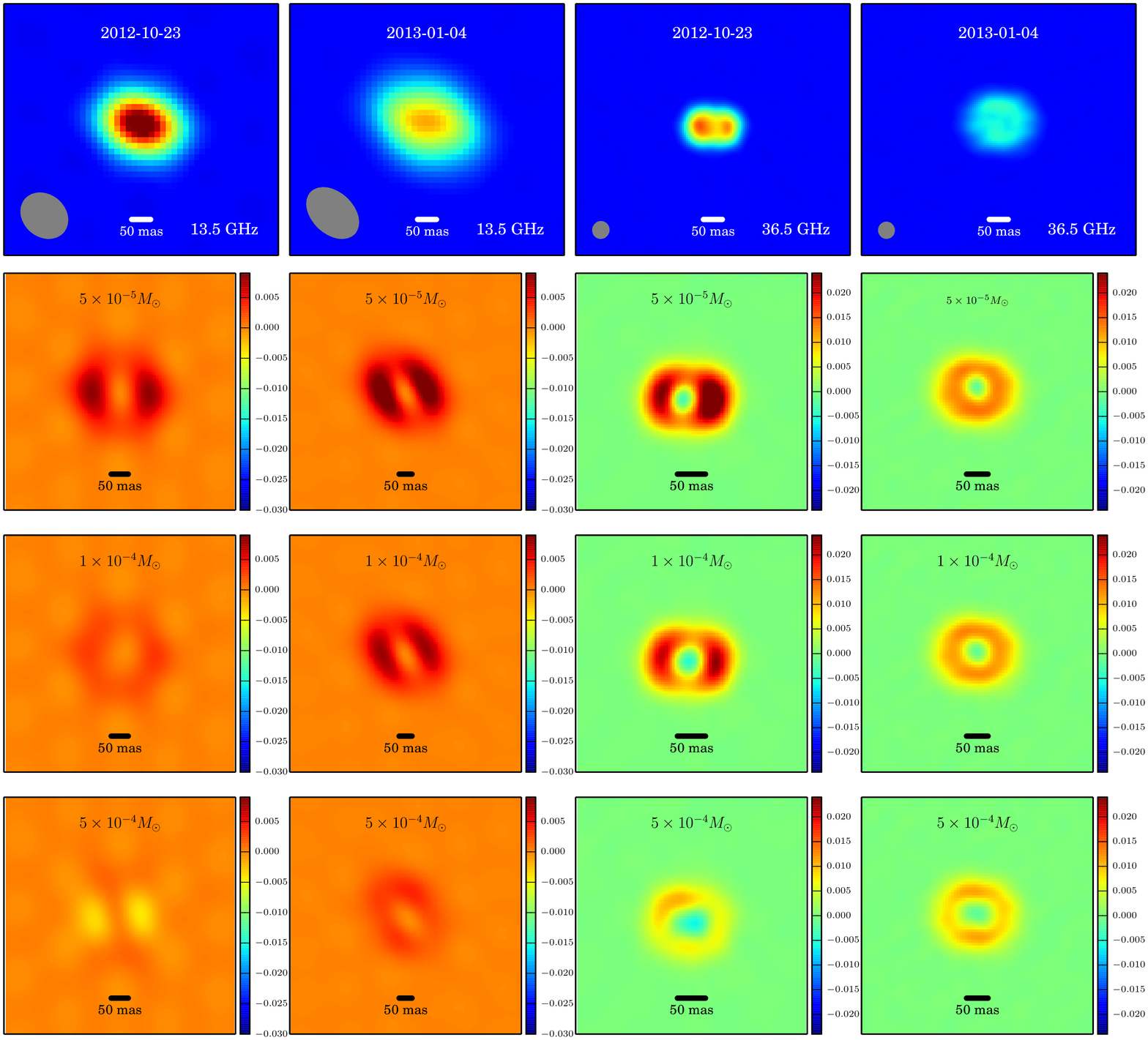}
\caption{Residual images (observed - model at 2 kpc) from 2012-10-23 (day 126) and 2013-01-04 (day 199).  Top row: VLA observation, with gray ellipse to indicate restoring beam.  Second row: Residual, SHAPE model $5\times10^{-5}M_{\odot}$.  Third row: $10^{-4}M_{\odot}$.  Bottom row: $5\times10^{-4}M_{\odot}$. A scalebar in the bottom center of each image indicate an angular size of 50 milliarcseconds.}
\label{res2kpc}
\end{figure*}
\end{center}

To compare the radio SHAPE models with observations, the models were convolved with the VLA restoring beam for the appropriate day and frequency.  Next, the models were scaled such that the peak flux density value in the model image matched the peak in the observed image.  This removes the effect of distance from the model images flux densities and allows for the direct comparison of structure between models and observations.  Residual images were created by subtracting the scaled models from the observed images.  Figures \ref{res1kpc} and \ref{res2kpc} show the residual maps with the radio SHAPE models at a distance of 1 kpc and 2 kpc, respectively.  The flux scale is the same for all residual images at a given frequency.  These images from the first A-configuration campaign represent a time when the nova ejecta were still largely optically thick.
From the residuals, it is apparent that the radio SHAPE models do a fair job of matching the observations at 13.5 GHz, but they do not match very well at 36.5 GHz.  The high frequency models become optically thin faster than the observations, especially in the east-west direction.  This is an indication that the ejecta are not perfectly smooth, but have some clumping.  Shore et al. (2013) indicated that a filling factor ($f$) of 0.1 was appropriate for this nova.  Such a filling factor would allow the high frequency diffuse ejecta to remain optically think longer and better match the observations due to the fact that optical depth scales as $\tau \propto f^{-0.5}$ (e.g., Heywood 2004, and Nelson et al. 2014).  

Also, note from Figure \ref{res2kpc} that all models with a distance of 2 kpc, except for the highest mass 13.5 GHz model for day 126, appear to be smaller than the observed images.  We therefore conclude that the true distance to the nova is likely less than 2 kpc.
\begin{center}
\begin{deluxetable}{ccccc}
\tablewidth{0 pt}
\tabletypesize{\small}
\tablecaption{ \label{tabrate}
Expansion Rates of V959 Mon}
\tablehead{ & & & Min. & Max. \\ Freq  & $\dot{\theta}$\tablenotemark{a} & Day 0\tablenotemark{b} & Distance\tablenotemark{c} & Distance\tablenotemark{d}\\
(GHz) & ($10^{-4}$ arcsec/day) & & (kpc) & (kpc)}
\startdata
Comp. 1\tablenotemark{e} & & & & \\
13.5 & $2.7\pm0.6$ & 8 & $1.0\pm0.3$ & $2.5\pm0.6$ \\
17.5 & $2.9\pm0.4$ & 15 & $1.0\pm0.2$ & $2.4\pm0.4$ \\
28.2 & $3.1\pm0.3$ & 25 & $0.9\pm0.1$ & $2.2\pm0.3$ \\
36.5 & $3.2\pm0.3$ & 25 & $0.9\pm0.1$ & $2.2\pm0.3$ \\
 & & & & \\
Comp. 2\tablenotemark{d} & & & & \\
13.5 & $3.2\pm0.6$ & 29 & $0.9\pm0.2$ & $2.2\pm0.5$ \\
17.5 & $3.2\pm0.3$ & 24 & $0.9\pm0.1$ & $2.2\pm0.4$ \\ 
28.2 & $3.4\pm0.2$ & 31 & $0.8\pm0.1$ & $2.0\pm0.3$ \\
36.5 & $3.4\pm0.2$ & 31 & $0.8\pm0.1$ & $2.0\pm0.3$ \\
 \enddata
\tablenotetext{a}{Angular expansion rate of the radius along the north-south axis.}
\tablenotetext{b}{The day, from \textit{Fermi} discovery, when the ejecta would have begun to expand according to linear fit.}
\tablenotetext{c}{Assuming an expansion velocity of $480\pm60$ km s$^{-1}$.}
\tablenotetext{d}{Assuming an expansion velocity of $1200\pm150$ km s$^{-1}$.}
\tablenotetext{e}{Component 1 refers to the westernmost component, Component 2 refers to the easternmost component}
\end{deluxetable} 
\end{center}
\section{Distance Derived from 2012-2013 Observations}

\begin{figure}
\includegraphics[width=3.8in]{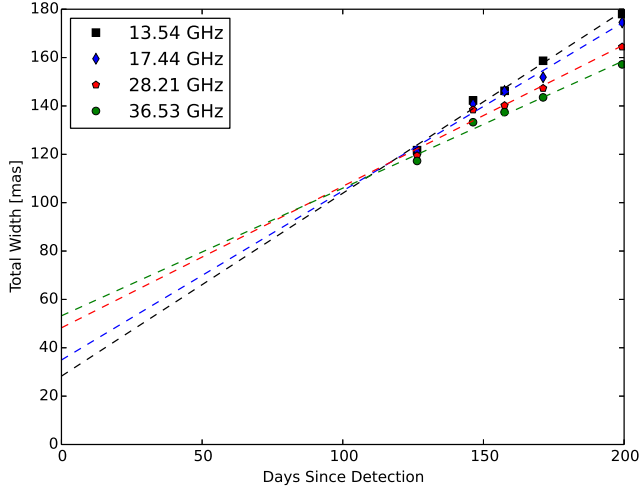}
\vspace{-0.45cm}
\caption{The growth of V959 Mon along the east-west axis (in units of milliarcsec) as a function of time. Measurements at the four distinct frequencies are plotted in different colors, as defined in the key at upper left. Linear fits are made to each frequency separately, and are plotted as colored dashed lines.}
\label{majorrate}
\end{figure}

\begin{figure}[t]
\includegraphics[width=3.8in]{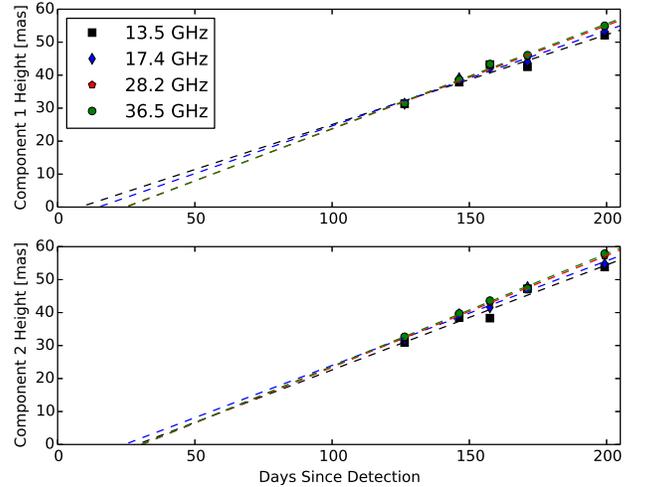}
\vspace{-0.45cm}
\caption{The growth of V959 Mon along the north-south axis (in units of milliarcsec) as a function of time. Top panel: western (left hand) component; bottom panel: eastern (right hand) component.  Measurements at the four distinct frequencies are plotted in different colors, as defined in the key at upper left of the top panel. Linear fits are made to each frequency separately, and are plotted as colored dashed lines.}
\label{minorrate}
\end{figure}

As discussed in Section 2, we performed Difmap model-fitting on the VLA observations of the nova in the \textit{(u,v)} plane with two elliptical Gaussian components for each of the five observations during the 2012-2013 VLA A-configuration campaign.  The Difmap model components show that the nova ejecta was expanding along both the east-west and north-south axes.  However, the material at the edges of the north-south axis remained optically thick throughout the 2012-2013 A-configuration campaign while the material along the east-west axis became optically thin at the edges.  We can see this as the Difmap model size along the east-west axis quickly became very dependent on observing frequency, but the north-south axis sizes remained consistent across all frequencies (see Figures~\ref{majorrate} and \ref{minorrate}).  

The angular extent of the ejecta in the north-south direction was found using the Difmap models.  First, a line was drawn between the centers of the two components.  Then, the angular extent was calculated as $0.5\times$ the total angular size perpendicular to this center line.  We then fit the growth of the calculated angular heights with time via linear least squares to determine the expansion rate.  The rates derived from the north-south axis fits are presented in Table~\ref{tabrate}.  Note that the derived rates do not very strongly with the observing frequency.  Also, the results for both components are similar despite the fact that they model independent areas of the ejecta.

In order to determine a distance to the nova V959 Mon with expansion parallax, we had to first determine the velocity of the ejecta.  This proved quite difficult for these observations due to the changing optical depth along the east-west axis.  We used the radio SHAPE models to estimate the appropriate velocities, but the changing optical depth led to significant uncertainties in exactly which material was being modelled at radio frequencies.  In order to address these uncertainties, we estimated the distance using two extremes in the possible velocity.  From the optical SHAPE models from \cite{Ribeiro13}, the squeeze of 0.8 puts a lower limit on the expansion velocity of $0.2\times$ the maximum velocity.  This gives us a minimum north-south velocity of 480 $\pm$ 60 km s$^{-1}$.  Also from the optical SHAPE models, because the maximum extent along the north-south axis was one-half the total extent along the east-west axis, the maximum velocity along the north-south axis is constrained to be $0.5\times$ the maximum velocity.  This gives a maximum north-south velocity of 1200 $\pm$ 150 km s$^{-1}$.  It is most likely that the true north-south velocity of the ejecta at the position of the Difmap models was somewhere in between these two extremes, but these still put strong limits on the distance.  

The distances were then calculated using
\begin{equation}
d = 5.78\times10^{-7}\ v/\dot{\theta}
\end{equation}
where $d$ is the distance in kpc, $v$ is the ejecta expansion velocity in km s$^{-1}$ (assumed to be constant), and $\dot{\theta}$ is the north-south angular expansion rate of the radius in arcseconds day$^{-1}$.  
The distances calculated from these velocity estimates are presented in Table~\ref{tabrate} and do not show a dependence on the observing frequency.  The mean distance using a velocity of 480 km s$^{-1}$ is $0.9\pm0.2$ kpc.  The mean distance using a velocity of 1200 km s$^{-1}$ is $2.2\pm0.4$ kpc.  These distances agree well with the estimates by \citep{Munari13}, but even the upper limit is significantly closer than what was used by Ackermann et al. (2014).

The rate of expansion in the north-south direction also provides weak evidence that the radio-emitting material did not begin its expansion until approximately a month after the first detection of the $\gamma$-rays.  Note from Figures~\ref{majorrate} and \ref{minorrate} that the fits to the expansion along both axes do not agree with expansion beginning at the time of discovery by \textit{Fermi}.  The fitted expansion rates for the east-west axis indicate that the expansion began before the nova was detected.  However, this is simply due to the fact that the edges of the ejecta become optically thin and the emitting surface recedes as it expands.  On the other hand, the fitted angular expansion rates for the north-south axis indicate that the expansion in the north-south direction did not begin until $\sim25$ days after detection (see Table~\ref{tabrate}).  Because the ejecta remains optically thick at the north-south edges, this delay may be real.  It is also possible that acceleration is important during early times.  However, given what is known about V959 Mon from Chomiuk et al. (2014b), it is also possible that this apparent stall is the time period when the binary is transferring energy to the nova envelope in a common-envelope-like event, culminating in the envelope gaining sufficient kinetic energy to escape the system (e.g., Livio et al 1990; Lloyd et al. 1997; Porter et al. 1998).  The recurrent nova T Pyx showed similar behavior during its outburst in 2011 in both radio \citep{Nelson14} and X-rays \citep{Chomiuk14a}.  It is also interesting to note that analysis of the X-ray light-curve of V959 Mon indicates the $t_{0}$ for the $n_{H}$ decline occurs around 30 days after detection (Nelson et al., in prep).  Because the $\gamma$-rays from V959 Mon were detected during the first 21.5 days of the nova evolution \citep{LAT14}, it seems likely that complex shock structures formed during this early interaction period.

\section{Distance Derived from 2014 Observations}

Chomiuk et al. (2014b) explain the evolution of nova V959 Mon as relating to two ejecta components.  The first component was denser, slower, and confined to the orbital plane of the binary system (the north-south direction).  The second component was a faster moving, less dense flow which was ``funnelled'' by the denser material and constrained to expand in the polar direction (east-west).  In the early stages of evolution, the fast polar flow remained optically thick and dominated the observed structure of the ejecta.  This fast polar flow was what we modelled with SHAPE (in both the optical and radio) in Sections 3 and 4.  An animation illustrating the evolution of optical depth in the fast component was presented in Figure 3 of the online version of Ribeiro et al. (2014)\footnote{http://iopscience.iop.org/0004-637X/792/1/57/article}.
The slower component only became obvious when the nova was imaged during the 2014 VLA A-configuration campaign.  As shown in Figure~\ref{Aconfig2}, particularly for the 13.5 GHz and 7.4 GHz images, the ejecta in the orbital plane remains visible and optically thick much longer than the fast material in the polar directions.  Because this material is associated with the 'pinched' region at the very center of the ejecta, its velocity must be the minimum derived from the optical SHAPE model.  That is, this material is moving at 480 km s$^{-1}$.  Therefore, the expansion parallax of these late time bright spots should provide the best estimate of the distance because we know the velocity to reasonably high certainty. 

Unfortunately, because the material is significantly dimmer at such late times, and because it has become so diffuse, it is harder to image and has more uncertainty in its position.  In fact, as mentioned in Section 2, the ejecta had become so diffuse that it was nearly completely resolved out at 28.2 and 36.5 GHz.  Also, we only had two epochs per observing frequency during the 2014 A-configuration, so we are not able to reliably determine the expansion rate.  Still, if we make some reasonable assumptions, we can calculate a most likely distance to the nova.

First, we assume that the material has been expanding since the nova was detected by \textit{Fermi}.  As mentioned in Section 3, the material may have started expanding about 25 days after expansion, but because these images were taken over 600 days after detection, an uncertainty of 25 days is only about 4\%, which is much less than the uncertainty in the north-south angular extent of the ejecta.  Second, we assume that the slow components have been expanding with a constant velocity of $480\pm120$ km s$^{-1}$.  The uncertainty in the velocity is twice as large as used in Section 4 in order to account for the fact that both the optical and radio SHAPE models did not expressly contain fast and slow components, but they nevertheless produce results similar to the observations (as demonstrated in Section 3).

Due to the diffuse nature of the ejecta, we were unable to determine the locations of the components by modelling in the \textit{(u,v)} plane with Difmap.  Instead, we used the AIPS task \verb|JMFIT| to determine the central location of the north and south components.  Once the angular separations between the two components are known, the distance can be calculated using
\begin{equation}
d = 1.155\times10^{-6}\ vt/\Delta\theta
\end{equation}
where $d$ is the distance in kpc, $v$ is the expansion velocity of the ejecta in km s$^{-1}$ with respect to the center of the nova, $t$ is the time from \textit{Fermi} detection in days, and $\Delta\theta$ is the separation between the two components in arcseconds.  The separations between the two components and the calculated distances are presented in Table~\ref{tabslow}.
The mean distance from these calculations is $1.4\pm0.4$ kpc.  This places the nova nicely in the middle of the lower and upper limits found in Section 4.  It also agrees well with distance estimates based on \ion{Na}{1}~D absorption (Munari et al. 2013).  

As an amusing side note, Chomiuk et al. (2012) used photometric parallax calculations to estimate the distance to V959 Mon based the $4.6$ GHz flux density on 2012 September 1 (Day 74), using 2012 June 22 as Day 0, and assuming spherical symmetry, an expansion velocity of $2000$ km s$^{-1}$, and a temperature of $10^{4}$K.  Every single one of these assumptions was incorrect.  The true Day 0 was 2012 June 19, so they assumed the ejecta had been expanding for 71 days instead of 74.  The ejecta is clearly \textit{not} spherical, which also means there is a large range in the expansion velocity.  Finally, the temperature estimate was off by a factor of approximately $6.5$.  Amazingly, despite all of these glaring inaccuracies, they estimated the distance to be $1.4$ kpc.  Purely by chance, the assumptions were all wrong in exactly the right way in order to result in the correct distance.  Clearly, such simplistic distance estimates should be treated with extreme caution.
\begin{center}
\begin{deluxetable}{cccc}[t]
\tablewidth{0 pt}
\tabletypesize{\footnotesize}
\setlength{\tabcolsep}{0.025in}
\tablecaption{ \label{tabslow}
V959 Mon Slowest Components}
\tablehead{ Freq  & Day\tablenotemark{a} & Separation & Distance\tablenotemark{b} \\
(GHz) &  & (arcsec) & (kpc)}
\startdata
13.5 & 615 & $0.242\pm0.02$ & $1.4\pm0.4$ \\
17.5 & 615 & $0.248\pm0.02$ & $1.4\pm0.4$ \\
7.4 & 676 & $0.212\pm0.04$ & $1.8\pm0.6$ \\
13.5 & 683 & $0.278\pm0.03$ & $1.3\pm0.4$ \\
17.5 & 683 & $0.302\pm0.03$ & $1.3\pm0.4$ \\ 
7.4 & 703 & $0.270\pm0.05$ & $1.4\pm0.4$ \\
\enddata
\tablenotetext{a}{The day, from \textit{Fermi} discovery.}
\tablenotetext{b}{Assuming an expansion velocity of $480\pm120$ km s$^{-1}$ and allowing for a 4\% uncertainty in the time since ejection.}
\end{deluxetable} 
\end{center}
To calculate the $\gamma$-ray luminosity of V959 Mon, Ackermann et al. (2014) used a distance of 3.6 kpc (based on UV spectral comparison to V1974 Cyg) which leads to a $\gamma$-ray luminosity of $3.7\times10^{35}$ erg s$^{-1}$.  Even using our absolute upper limit (from Section 4) of 2.6 kpc, the luminosity would be $1.4\times10^{35}$ erg s$^{-1}$.  Using our most probable distance of 1.4 kpc, we find the $\gamma$-ray luminosity is more likely to be approximately $0.6\times10^{35}$ erg s$^{-1}$.  This is much less than the luminosity for the other $\gamma$-ray-detected novae, the lowest being V339 Del at $2.6\times10^{35}$ erg s$^{-1}$ and the highest being V1324 Sco at $8.6\times10^{35}$ erg s$^{-1}$ \citep{LAT14}.  However, it must be noted that the distances to these novae remain uncertain.  However, we suspect that both V339 Del and V1324 Sco have larger distances than reported in \cite{LAT14}, leading to even larger $\gamma$-ray luminosities for these objects.  In fact, using the distance to V339 Del of $4.54$ kpc as reported by Schaefer et al. (2014), the $\gamma$-ray luminosity is closer to $3.0\times10^{35}$ erg s$^{-1}$.  Work by Finzell et al. (in prep) also indicates that V1324 Sco is likely further away, and thus more luminous, than reported in \cite{LAT14}.  This makes V959 Mon even more of a low-luminosity outlier than it already appears to be.  

\section{Conclusions}

Using expansion parallax techniques and high-resolution radio imaging, we find that the distance to nova V959 Mon is between $0.9\pm0.2$ kpc and $2.2\pm0.4$ kpc, with a most probable distance of $1.4\pm0.4$ kpc.  These results agree well with estimates of the distance based on \ion{Na}{1}~D absorption \citep{Munari13}.  Sophisticated 3-D radio models created with SHAPE based on optical spectroscopy constrain the expansion velocity of the ejecta.  Comparison of the VLA images with the synthetic radio SHAPE images also indicate the nova is closer than 2.0 kpc.

Tracking the expansion of the ejecta in the radio images also indicates that the radio-emitting material may not have begun expanding until roughly 25 days after the nova was discovered by \textit{Fermi}.  This apparent ``stall'' time is suspiciously similar to the time when $n_{H}$ decline occurs in the X-ray observations ($\sim$30 days; Nelson et al., in prep), and the total time that V959 Mon was detected by \textit{Fermi} (21.5 days; Ackermann et al. 2014).  As proposed in \cite{Chomiuk14b}, it is likely that during the first 25 days the nova was in a common-envelope-like phase where complex shock structures formed and led to both synchrotron and $\gamma$-ray emission.  Interestingly, the infrared interferometric observations of another \textit{Fermi}-detected nova, V339 Del, from Schaefer et al. (2014) also shows a sudden increase in expansion rate around 30 days after initial detection.  It is possible that many (and perhaps all) novae undergo this common-envelope-like phase during the first month of their evolution.

The morphology of other \textit{Fermi}-detected classical novae (i.e., novae without a giant companion) are still largely unknown (e.g., Schaefer et al. 2014; Ackermann et al. 2014).  Further modelling and observations are needed to investigate if they are mostly bipolar in shape like V959 Mon, or perhaps a simpler spherical shape.  However, in the case of a spherical shape, it is difficult to explain where the $\gamma$-rays would originate without a multi-phase ejection model that produces internal shocks.  Also, if the common-envelope-like phase is as ubiquitous as we suspect, nearly all classical novae should have complicated bipolar shapes (e.g., Lloyd et al. 1997; Porter et al. 1998).  Any apparent spherically-shaped nova may simply be due to the orientation with respect to our line-of-sight.

Using our calculated distance to V959 Mon of 1.4 kpc, the $\gamma$-ray luminosity is approximately $0.6\times10^{35}$ erg s$^{-1}$.  This is much lower than the other three \textit{Fermi}-detected novae presented in \cite{LAT14}, indicating that the range of $\gamma$-ray luminosities for novae is at least a factor of 10.  If V959 Mon had been at a more typical distance for a nova (e.g., between 4 and 8 kpc), it probably would not have been detected by \textit{Fermi}.  Furthermore, the existence of one such low $\gamma$-ray luminosity nova indicates that there are likely many other novae with $\gamma$-ray emission that are undetectable simply due to their distance.  In fact, it is possible that all novae emit $\gamma$-rays, but we only detect the most luminous and/or nearest of them.

The difference in the $\gamma$-ray luminosity between V407 Cyg and V959 Mon are likely due to the presence of wind from a giant companion in V407 Cyg while V959 Mon has a main sequence companion with no such wind.  The differences in luminosity between V959 Mon and V339 Del or V1324 Sco require a different explanation.  However, it is well-known that novae have a large range in ejected mass and the ejecta velocities and these factors are likely to be important in determining their $\gamma$-ray luminosities.  It should also be noted that the distances to the other \textit{Fermi}-detected novae remain highly uncertain, and we suspect they are even more luminous than reported in \cite{LAT14}.  We look forward to future work in determining better the distances to these and future novae, both with the techniques presented here and with the Gaia\footnote{http://sci.esa.int/gaia/} space telescope currently beginning its mission operations.

\acknowledgements
The authors thank the anonymous referee for their constructive (and extremely pleasant) criticism.
The authors thank W. Steffen and N. Koning for useful discussions on the use of SHAPE, and T. Finzell for useful discussions on V1324 Sco.
J.~L. and L.~C. were supported in part by NASA Fermi Guest Investigator grant NNH13ZDA001N-FERMI.
V.~A.~R.~M.~R. was supported in part by the Radboud Excellence Initiative and the South Africa SKA Project.
T.~N. was supported in part by NASA award NNX13A091G.
J.~L.~S. and J.~W. were funded in part by NSF award AST-1211778.
The National Radio Astronomy Observatory is a facility of the National Science Foundation operated under cooperative agreement by Associated Universities, Inc. This research made use of APLpy, an open-source plotting package for Python hosted at http://aplpy.github.com.\\

{\it Facilities:} \facility{Karl G. Jansky VLA}

\end{document}